\documentclass[aps,prl,epsfigure,twocolumn,superscriptaddress]{revtex4-2}
\usepackage[colorlinks=true,linkcolor=blue,urlcolor=blue,citecolor=blue,pdfusetitle]{hyperref}
\usepackage[utf8]{inputenc}
\usepackage[english]{babel}
\usepackage{amsmath}
\usepackage[caption = false]{subfig}
\usepackage{graphicx,epstopdf}
\usepackage{blindtext}
\usepackage[table,xcdraw]{xcolor}
\usepackage{lipsum}
\usepackage{amsfonts}
\usepackage{bbold}
\usepackage{bbm}
\usepackage{amssymb}
\usepackage{enumerate}
\usepackage{color}
\usepackage{xcolor}
\usepackage{latexsym}
\usepackage{physics}
\usepackage{times,txfonts}
\usepackage[normalem]{ulem}

\newcommand{\Ocal}{\mathcal{O}}

\newcommand{\1}{\mathbbm{1}}

\newcommand{\SubFig}[2]{\ref{#1}{\color{blue}#2}}

\definecolor{MyGreen}{RGB}{0, 179, 134}
\definecolor{MyRed}{RGB}{255, 102, 102}

\newcommand{\UFSCar}{Departamento de Física, Universidade Federal de São Carlos, \\Rodovia Washington Luís, km 235 - SP-310, 13565-905 São Carlos, SP, Brazil}

\newcommand{\Nice}{Universit\'e C\^ote d'Azur, CNRS, Institut de Physique de Nice, 06560 Valbonne, France}

\usepackage{orcidlink}

\begin{document}


\title{Collateral coupling between superconducting resonators: Fast and high fidelity \\generation of qudit-qudit entanglement}

\author{Pedro Rosario}
\email{pedrorosario@estudante.ufscar.br }
\affiliation{\UFSCar}

\author{Alan C. Santos~\orcidlink{0000-0002-6989-7958}}
\email{ac\_santos@df.ufscar.br}
\affiliation{\UFSCar}

\author{C. J. Villas-Boas~\orcidlink{0000-0001-5622-786X}}
\email{celsovb@df.ufscar.br}
\affiliation{\UFSCar}

\author{R. Bachelard~\orcidlink{0000-0002-6026-509X}} 
\email{romain@ufscar.br}
\affiliation{\UFSCar}
\affiliation{\Nice}


\begin{abstract}
Superconducting circuits are highly controllable platforms to manipulate quantum states, which make them particularly promising for quantum information processing. We here show how the existence of a distance-independent interaction between microwave resonators coupled capacitively through a qubit offers a new control parameter toward this goal. This interaction is able to induce an idling point between resonant resonators, and its state-dependent nature allows one to control the flow of information between the resonators. The advantage of this scheme over previous one is demonstrated through the generation of high-fidelity NOON states between the resonators, with a lower number of operations than previous schemes. Beyond superconducting circuits, our proposal could also apply to atomic lattices with clock transitions in optical cavities, for example.
\end{abstract}

\maketitle


\emph{Introduction.---} Superconducting integrated circuits~\cite{Blais_2004,Koch_2007,Rigetti_2010,Blais_2021_re} are currently one of the most promising platforms for building quantum computers, due to their potential for scalability, multi-qubit tunable interaction and high fidelity processing of quantum information ~\cite{JMartinis_2019_Sy,Yulin_2021,Zhang_2022}. Differently from cold atoms or trapped ions, the coupling between the components of a circuit occurs between nearest neighbour, although the physical proximity between the different components can result into additional (parasitic) couplings, generating cross-talks and affecting the fidelity of protocols~\cite{Peng_2022,Xu:20}. The advantage of nearest-neighbour interactions is that a local transfer of information can be operated, while the rest of the circuit remains unaffected, using for example a suitable detuning between the qubits or local oscillating fields~\cite{Sarah2016,Mates2019,Nuerbolati2022,Dai2022,Tripathi2022,Berke2022}. 


Of particular interest are systems of superconducting resonators, due to their high performance to mediate interaction between several qubits in integrated quantum circuits~\cite{xiang_2013,Omran:19,Mirhosseini:19,Wang:20,Marinelli:23}. Furthermore, when using superconducting qubit-resonators hybrid systems to process quantum information, computation beyond qubits can be performed, where arbitrary quantum states of \textit{qudits} can be generated~\cite{Hofheinz_2008,Hofheinz_2009}. The configuration of two resonators coupled through an intermediate qubit has proven especially successful, with theoretical and experimental works demonstrating the generation of entangled photon Fock states~\cite{Mariantoni_2008,HWang_2011}, two-mode cat states of electromagnetic fields~\cite{CWang_2016}, and for the improvement of quantum non-demolition measurements in the Fock basis in a resonator~\cite{BRJohnson_2010}.

\begin{figure}[t!]
	\includegraphics[width=0.9\columnwidth]{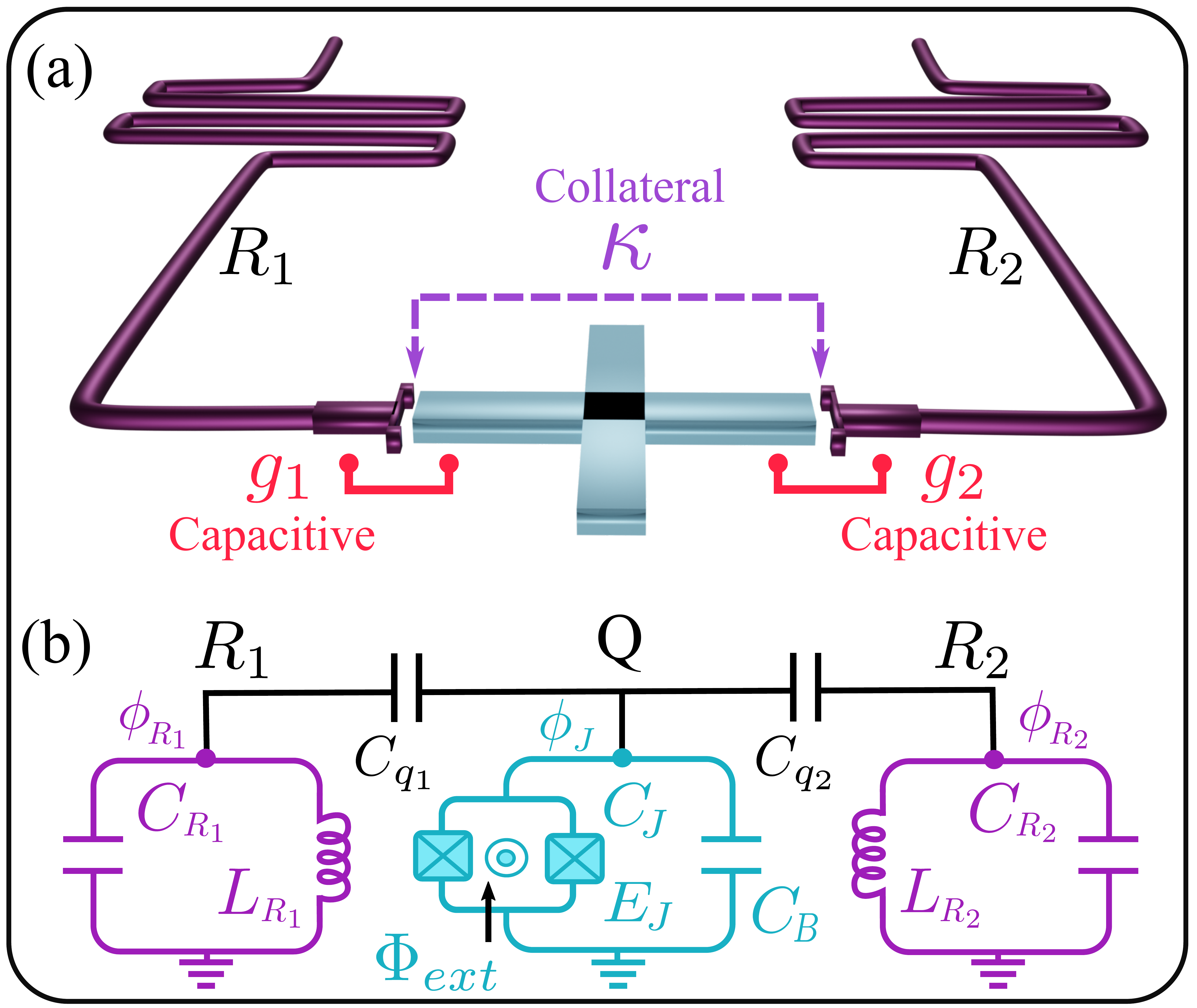}
	\caption{ (a) Two superconducting resonators ($R_{1}$, $R_{2}$) coupled to a transmon (Q) present a  qubit-mediated interaction ($\kappa$) due to the capacitance network of the system and direct atom-resonators capacitive couplings ($g_{k}$). The setup is equivalent to an atom inside two crossed optical cavities . (b) Lumped superconducting circuit showing the capacitance and inductance networks.}
	\label{Fig:Scheme}
\end{figure}

We here discuss how the mere fact of using a qubit to transfer information from one resonator to the other actually induces a new coupling between the resonators, hereafter called ``collateral'' coupling. Different from a mediated interaction, which is obtained from tracing over the variables of the mediating component (here the qubit), it stems from switching from the Lagrangian description of the integrated circuit, where only nearest-neighbour coupling terms are present, to its Hamiltonian representation~\cite{Garcia_Ripoll2022-go,Rasmussen:21}. Interaction of this nature also is present in system of tunable coupling between superconducting artificial atoms~\cite{Yan:18}, and for systems of atoms in optical cavities, as we shall see.
In the resonator-qubit-resonator sketched in Fig.~\ref{Fig:Scheme}, a distance-independent interaction between the two resonators emerges, which scales as the product of the two direct capacitive resonator-qubit coupling. 


Although the collateral term is a correction to the overall Hamiltonian, it represents a substantial contribution in the case of superconducting qubits, able to generate an idling point for the resonators dynamics. 
The dependence of the idling point on the qubit state provides a new control parameter to engineer quantum states, of which we take advantage to generate efficiently NOON states between the resonators. Finally, we discuss how the collateral coupling term could turn equally important for other platforms such as cold atoms with clock transitions in crossed optical cavities.




\emph{Modelling of the superconducting circuit.---} Let us consider a linear circuit composed of two resonators $R_{1}$ and $R_2$ each coupled to an intermediate qubit Q (qubit-resonator capacitance $C_{q_{1,2}}$), see Fig.~\SubFig{Fig:Scheme}{(b)}. There is no direct interaction between the resonators, a situation encountered when the distance between them is large enough to prevent parasitic coupling.
The qubit is a frequency-tunable transmon~{\cite{barends2013}}, whose frequency $\omega(\Phi_{\mathrm{ext}})$ can be tuned through the external flux $\Phi_{\mathrm{ext}}$, being its natural frequency $\omega_{0} = \omega(\Phi_{\mathrm{ext}}=0)$. We assume a transmon composed of a superconducting loop with two identical Josephson Junction with energy $E_J$ and capacitance $C_{J}$, shunted by an external capacitor $C_{\mathrm{B}}$. As for the two resonators, they are characterized by their internal capacitance $C_{R_{n}}$ and inductance $L_{R_{n}}$, with $n=1,\ 2$.

The equations of motion are derived by first writing down the system Lagrangian. Applying Kirchhoff’s laws and the method of nodes~\cite{Vool:17,Rasmussen:21} to the circuit diagram presented in  Fig.~\SubFig{Fig:Scheme}{(b)}, we obtain the following Lagrangian:
\begin{eqnarray}
L(\vec{\phi},\vec{\dot{\phi}})&=(C_{\mathrm{T}}+C_{q_{1}}+C_{q_{2}})\frac{\dot{\phi}^{2}_{J}}{2} + \sum_{n=1}^{2}(C_{R_{n}}+C_{q_{n}})\frac{\dot{\phi}^{2}_{R_{n}}}{2}\nonumber\\
	&- C_{q_{1}}\dot{\phi}_{J}\dot{\phi}_{R_{1}}-C_{q_{2}}\dot{\phi}_{J}\dot{\phi}_{R_{2}}- U(\vec{\phi}),
 \label{eq:Lag}
\end{eqnarray}
where $\vec{\phi} = (\phi_{J},\phi_{R_{1}},\phi_{R_{2}})$, with $\phi_{J}$ and $\phi_{R_{n}}$ the node flux of the qubit and of the $n$-th resonator, respectively, while $C_{\mathrm{T}}=C_{J} + C_{\mathrm{B}}$ is the total qubit box capacitance. The potential term $U(\vec{\phi})$ is given by 
$U(\vec{\phi})=\phi^{2}_{R_{1}}/2L_{R_{1}} +\phi^{2}_{R_{2}}/2L_{R_{2}}-\tilde{E}_{J}\cos(\phi_{J} / \Phi_{0}),$
with $\tilde{E}_{J}= 2 E_{J} \cos(\Phi_{\mathrm{ext}}/2\Phi_{0})$ the Josephson energy, and $\Phi_{0}=\hbar/2e$ the quantum of magnetic flux through the qubit. Switching to the Hamiltonian formalism leads to:
\begin{align}
	H&\approx\frac{q^{2}_{J}}{2C_{\mathrm{T}}}- \tilde{E}_{J}\cos(\frac{\phi_{J}}{\Phi_{0}})
	+ \frac{q^{2}_{R_{1}}}{2C_{R_{1}}} + \frac{\phi^{2}_{R_{1}}}{2L_{R_{1}}} 
	+ \frac{q^{2}_{R_{2}}}{2C_{R_{2}}} + \frac{\phi^{2}_{R_{2}}}{2L_{R_{2}}} \nonumber \\
	&+ \frac{C_{q_{1}}}{C_{R_{1}}C_{\mathrm{T}}}q_{J}q_{R_{1}}+ \frac{C_{q_{2}}}{C_{R_{2}}C_{\mathrm{T}}}q_{J}q_{R_{2}} + \frac{C_{q_{1}}C_{q_{2}}}{C_{R_{1}}C_{R_{2}}C_{\mathrm{T}}}q_{R_{1}}q_{R_{2}} ,
 \label{eq:H_classical}
\end{align}
where we have introduced the momentum $q_{k} = \partial L(\vec{\phi},\vec{\dot{\phi}}) / \partial \mathbin{\textcolor{black}{\dot{\phi_{k}}}}$, conjugate to the node flux $\phi_k$. Hamiltonian \eqref{eq:H_classical} has been derived in the regime where $C_{\mathrm{T}}= C_{J} + C_{\mathrm{B}} \gg C_{q_{k}}$ and $C_{R_{k}} \gg C_{q_{k}}$~\cite{SupInf}, where corrections of the order $\Ocal(C_{q_{k}}/C_{T})$ and $\Ocal(C_{q_{k}}/C_{R_{k}})$ are neglected.


The last term in Eq.~\eqref{eq:H_classical} is of particular interest, since it corresponds to a direct coupling between the resonators, due to the presence of a qubit coupled to each of them. We call it ``collateral'' capacitive coupling: It does not correspond to a ``mediated'' or ``effective'' interaction, since we did not trace out over the qubit degree of freedom, neither did we operate a specific transformation or operation on the variables of the Hamiltonian that would make it appear (the approximation on the capacities is a mere convenience, and does not affect the existence of the collateral term). Furthermore, it cannot be considered as a parasitic coupling, which refers to an additional coupling typically due to the physical proximity of the different components~\cite{Yang:12}. 
Indeed, the collateral term does not depend on the distance between the resonators, but only to the capacitive coupling of each with the qubit, such that it is the dominant interaction in the limit in which parasitic couplings vanish~\cite{SupInf}. As we discuss later, it is in essence similar to a mode-mode coupling term already derived by Fermi in the context of a charged particle interacting with electromagnetic modes~\cite{Fermi_32}. 


We then proceed by quantizing the circuit Hamiltonian, adopting the standard approach of the second quantization~\cite{Vool:17} and introducing the creation and annihilation operators for the resonators, $\hat{a}_{k}^{\dagger}$ and $\hat{a}_{k}$, as well as for the qubit, $\hat{\sigma}^{+}$ and $\hat{\sigma}^{-}$. This leads to the quantized Hamiltonian~\cite{SupInf}:
\begin{align}
	\hat{H}&=\hbar\omega_q \left(\hat{\sigma}^{+}\hat{\sigma}^{-}+\frac{\1}{2}\right)+\hbar\sum_{k=1}^{2}\omega_{R_{k}}\left(\hat{a}_{k}^{\dagger}\hat{a}_{k}+\frac{\1}{2}\right)\nonumber\\
	&+\hbar\sum_{k=1}^{2} g_{k}(\hat{a}_{k}^{\dagger}\hat{\sigma}^{-}+\hat{a}_{k}\hat{\sigma}^{+}) + \hbar\kappa(\hat{a}_{1}^{\dagger}\hat{a}_{2}+\hat{a}_{1}\hat{a}_{2}^{\dagger}) , \label{Eq:Hfull}
\end{align} 
where $\hbar \omega_q = \sqrt{8E_C \mathbin{\textcolor{black}{\tilde{E}_{J}}}}-E_C$ is the qubit frequency, and $\omega_{R_{k}}=1 / \sqrt{C_{R_{k}} L_{R_{k}}}$ the resonators ones. The Hamiltonian is valid when $\min \{\omega_q,\omega_{R_{k}}\} \gg \max \{g_{k},\kappa \}$ after applying the rotating-wave approximation, where the resonator-qubit and resonator-resonator coupling strengths read
\begin{subequations}
\label{Eq:Parameters}
\begin{align}
	g_{k} &= - \frac{1}{\hbar } \frac{C_{q_{k}}}{\sqrt{ C_{\mathrm{T}} C_{R_{k}}}} \left(2E_C\tilde{E}_{J}\epsilon_{R_{k}}^2\right)^{1/4},\label{eq:q_circuit} \\
	\kappa &= \frac{C_{q_{1}}C_{q_{2}}}{C_{\mathrm{T}}\sqrt{C_{R_{1}}C_{R_{2}}}}\frac{\sqrt{\epsilon_{R_{2}}\epsilon_{R_{1}}}}{\hbar} \label{eq:k_circuit},
\end{align}
\end{subequations}
with $E_C=e^{2}/2C_{\mathrm{T}}$ the qubit charging energy and $\epsilon_{R_{k}} = \hbar \omega_{R_{k}}/2$ the zero-point energy of the $k$-th resonator. 

The collateral coupling strength $\kappa$ depends on the different parameters of the electronic elements of the superconducting chip, since the Josephson junction capacitance is encoded in $C_{\mathrm{T}}$. While it does not depend on the physical distance between the two resonators, it scales as $C_{q_{1}}C_{q_{2}}$, and acts as a correction to the dynamics composed of the direct resonator-qubit coupling $C_{q_{k}}$ only. As for their relative amplitude, let us consider the following parameters for the capacitances and inductances~\cite{Vool:17}: $C_{R_{k}} = 4C_{\mathrm{T}} = 400.0$ fF and $L_{R_{k}} = 0.8$~nH correspond to a frequency $\omega_{R_{k}}/2\pi\approx 8.9$~GHz, coupling capacitances of $C_{q_{k}} = C_{R_{k}} / 50  \approx 8.0$~fF and a Josephson energy $E_{J}= 50 E_C$. This leads to the coupling values $g_{k}/2\pi\approx -139.7$~MHz and $\kappa/2\pi \approx 7.1$~MHz, so the ratio between the collateral resonator-resonator coupling and the direct resonator-qubit coupling is of the order $\kappa/g_{k} \sim 20$. Even without a detailed analysis of the dynamics, such a correction cannot be neglected in the context of high-fidelity on-resonance operations, where the last term in Eq.~\ref{Eq:Hfull} becomes relevant. But more importantly, as we shall now see, it can even induce an idling point for the resonator dynamics. 


\begin{figure}[t!]
\includegraphics[width=\columnwidth]{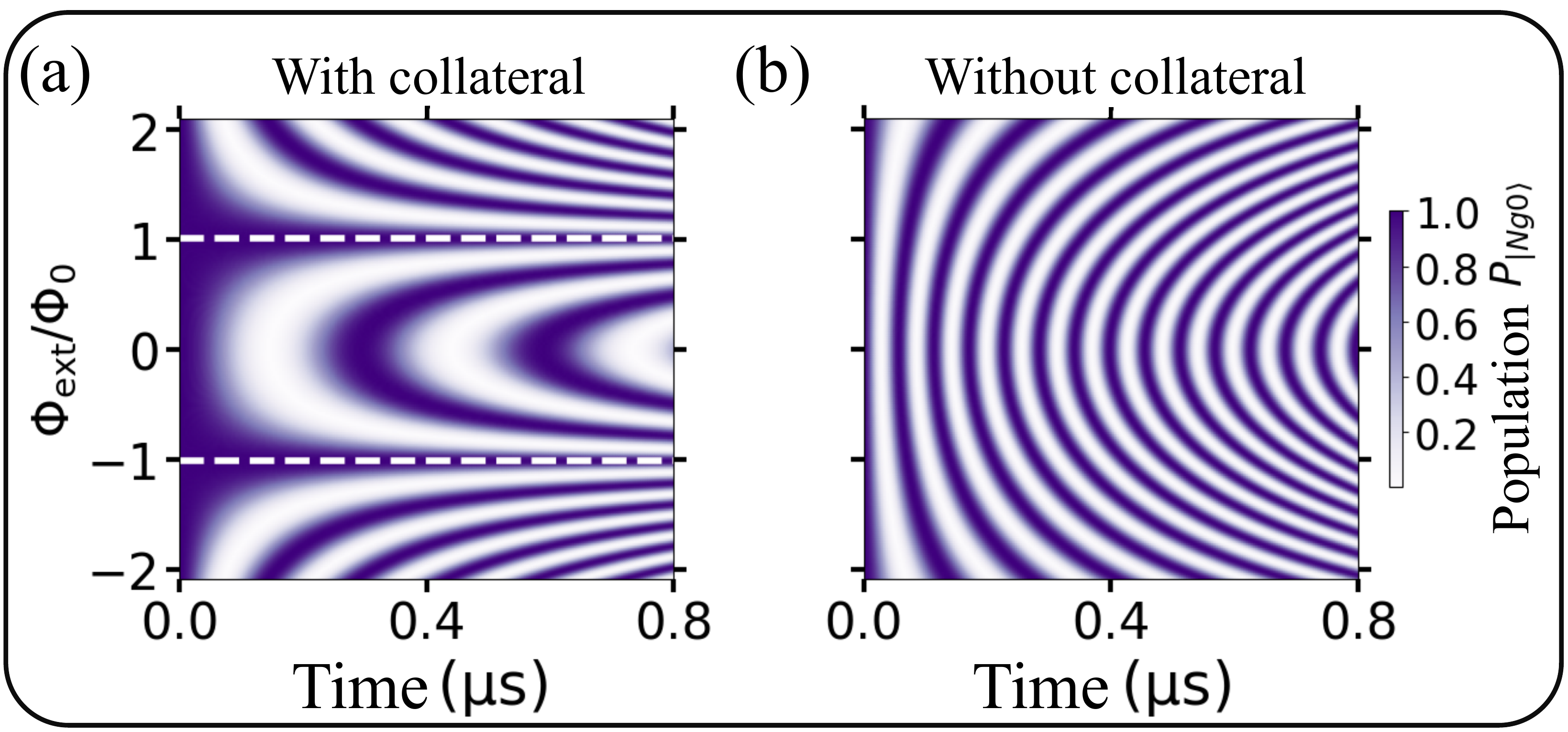}
\caption{Dynamics of the population $P(t)=|\braket{Ng0}{\psi(t)}|^{2}$ of state $\ket{Ng0}=\ket{N}_{R_{1}}\ket{g}_{\text{Q}}\ket{0}_{R_{2}}$ (considering that the system is initially in that state), as a function of time and the qubit external flux (a) simulating the full Hamiltonian~\eqref{Eq:Hfull} with idling point at $\Phi_{\mathrm{ext}}/\Phi_{0}\approx 1.013$, and (b) cancelling artificially the collateral term ($\kappa=0$, as it is usually assumed~\cite{Strauch:10,Su_2014,HWang_2011}). Simulations realized with $C_{R_{k}} = 4C_{\mathrm{T}} = 400.0$ fF, $L_{R_{k}} = 0.8$~nH, $C_{q_{k}} = C_{R_{k}} / 50 $ and $E_{J}= 70 E_C$.} 
	\label{Fig:chevron}
\end{figure}

{\em On-resonance idling point.---} The effect of the collateral coupling $\kappa$ on the circuit dynamics can be appreciated in the population evolution for state $\ket{Ng0}=\ket{N}_{R_{1}}\ket{g}_{Q}\ket{0}_{R_{2}}$ presented in Fig.~\ref{Fig:chevron}. After being initialized in that state, the resonators are put at the same frequency, and the excitation starts oscillating back and forth between resonator $R_1$ and $R_2$, at a frequency which characterizes the resonator-resonator effective coupling strength $g^{\mathrm{eff}}_{R_{1}R_{2}}$~\cite{Caldwell:18,Casparis:18,Leonard:19,KimEunjong:21,Bruno:22}. As it can be seen from comparing the dynamics with and without collateral term (panels (a) and (b), respectively), this new contribution alters dramatically the resonator-resonator coupling. 


For the set of parameters considered in Fig.~\ref{Fig:chevron}, the resonator-resonator effective coupling between resonators cancels for $\Phi_{\mathrm{ext}}\approx\pm \Phi_0$, and an idling point emerges: The population remains blocked in state $\ket{Ng0}$ (dashed white lines in Fig.~\SubFig{Fig:chevron}(a)). This corresponds to a cutoff frequency for the qubit, at which the transfer of information between the resonators is prevented. The effect of the collateral coupling on the circuit dynamics is thus particularly striking in this regime, despite its small relative weight as compared to the direct resonator-qubit coupling ($g_{k}/\kappa \sim 20$). As we shall see now, this stems from the fact that while the collateral coupling depends on $\kappa$ only, the effective interaction between the resonators can be tuned independently through the different circuit parameters, so the two interactions can cancel.

The emergence of this idling point is more easily understood from the effective dynamics. Let us now focus on the case where resonators have the same frequency, $\omega_{R_{1}}=\omega_{R_{2}}=\omega_{R}$. We enter the  dispersive regime for the qubit-mediated interaction by choosing a large detuning $\Delta = \omega_{q} - \omega_{R}$, that is, $|\Delta| \gg g_{1,2}$. Since the qubit remains in its ground state for all the evolution, and assuming the rotating wave approximation, we obtain the effective Hamiltonian, now tracing over the qubit degree of freedom~\cite{SupInf} 
\begin{equation}
H_{\mathrm{eff}}=
 \hbar\sum_{n=1}^{2}\eta_{n}\hat{a}_{n}^{\dagger}\hat{a}_{n}+ \hbar g^{\mathrm{eff}}_{R_{1}R_{2}} \left(\hat{a}_{1}\hat{a}_{2}^{\dagger}+\hat{a}_{1}^{\dagger}\hat{a}_{2}\right), 	\label{Eq:Heff}
\end{equation}
with $\eta_{n} = g^{2}_{n}/\Delta$, and where 
the first and second terms stand for the free Hamiltonian, which describes the energy shifts, but no population exchange. The third term describes the resonator-resonator coupling, $g^{\mathrm{eff}}_{R_{1}R_{2}}=g_{1}g_{2}/\Delta+\kappa$: It is a combination of the collateral term $\kappa$ with the qubit mediated interaction, characterized by the coupling $g_{1}g_{2}/\Delta$. The idling point corresponds to the situation where these two contributions cancel, $\Delta=-g_{1}g_{2}/\kappa$ in the dispersive regime discussed here. In that case, only the free Hamiltonian remains, which does not allow for information transfer between the resonators. In terms of external flux, this condition translates into~\cite{SupInf}
\begin{equation}
\Phi_{\mathrm{ext}}/\Phi_{0}=2\arccos{\left(\frac{\left(\kappa C_{R_{k}}C_{T}(E_{C}+\hbar \omega_{R_{k}})\right)^{2}}{E_{C}E_{J}(4\kappa  C_{R_{k}}C_{T}+C^{2}_{q_{k}}\omega_{R_{k}})^{2}}\right)}.
    \label{Eq:Idle_Flux}
\end{equation}

Note that the existence of the idling point relies on a finite value of the collateral term $\kappa$. Indeed, if $\kappa=0$, the idling point condition~\eqref{Eq:Idle_Flux} imposes $\Phi_{\mathrm{ext}}/\Phi_{0}=\pi(2m+1)$ for the external flux, with $m\in \mathbb{N}$. This is satisfied when $\tilde{E}_{J}=0$, and consequently $g_{k}=0$, which represents a trivial solution where qubit and resonators do not interact at all. On the other hand,  
the effective Hamiltonian approach results in $\Phi_{\mathrm{ext}}/\Phi_{0}\approx 1.015$, in excellent agreement with the simulations of the total Hamiltonian presented in Fig.~\SubFig{Fig:chevron}{a}.  In context of coupled qubit systems, such an idling point has been experimentally detected in scenarios where the parasitic interaction is present~\cite{Li:20,Bruno:22}.


\emph{NOON states.---} The idling point resulting from the collateral term $\kappa$ in turn allows for the generation of specific states, such as NOON states. These two-mode highly entangled state of the form $\ket{{\mathrm{NOON}}}=\frac{1}{\sqrt{2}}(\ket{Ng0}+\ket{0gN})$ are of interest for testing Bell-type inequalities~\cite{Wildfeuer_2007} and bring new opportunities to explore quantum communication protocols~\cite{JPan_2012}. Their generation in superconducting qubits has been investigated~\cite{Strauch:10,Merkel_10,Sharma:16}, yet the absence of idling point or state-selective interaction makes the process challenging.

\begin{figure}[t!]
	\includegraphics[width=\columnwidth]{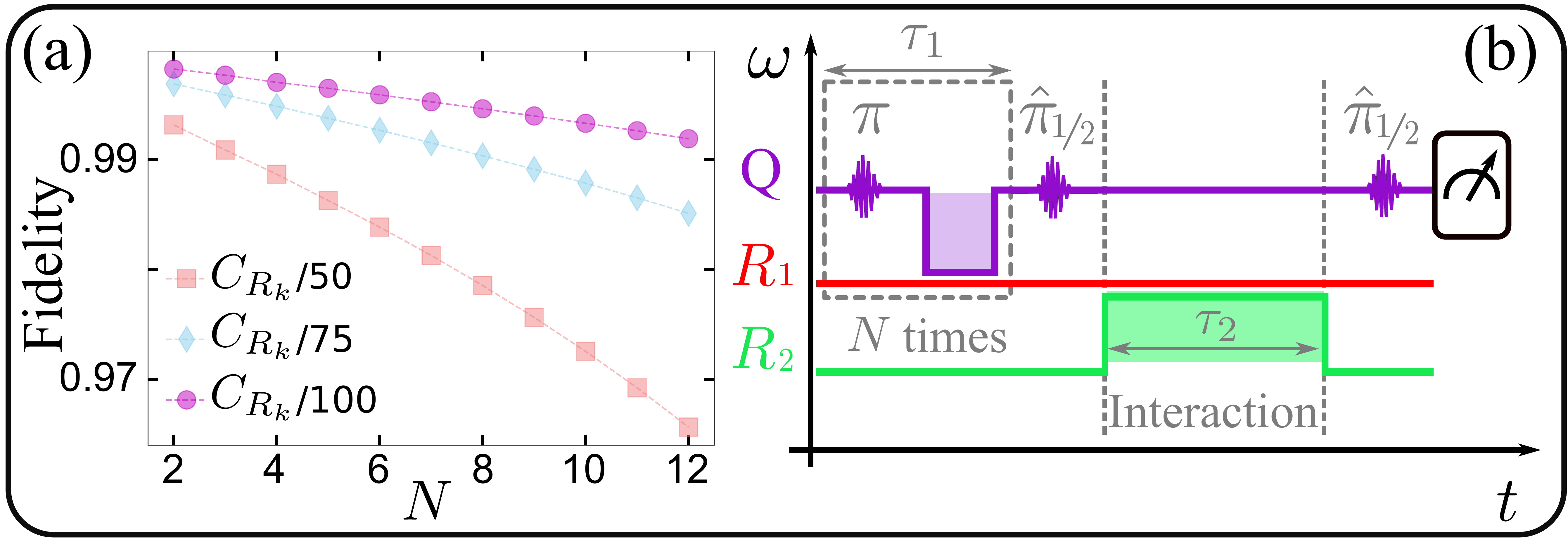}
	\caption{(a) Fidelity as function of the number of excitations $N$. The fidelity increase when the effective limit $C_{q_{k}}\ll C_{R_{k}}$  is increasingly satisfied. (b) Pulse sequence to generate the NOON states following our protocol. The parameters used are $C_{R_{k}} = 4C_{\mathrm{T}} = 400.0$ fF, $L_{R_{k}} = 0.8$~nH and $E_{J}= 70 E_C$, considering the cases $C_{q_{k}} = C_{R_{k}} / 50 $, $C_{q_{k}} = C_{R_{k}} / 75 $ and $C_{q_{k}} = C_{R_{k}} / 100 $ respectively.}
	\label{Fig:NooNF}
\end{figure}

Let us now describe the protocol to generate efficiently NOON states based on the idling point. First we prepare the system in state $\ket{Ng0}$, by putting qubit $R_{1}$ on resonance with qubit $\text{Q}$ (but not with $R_{2}$), so that the excitation created in the atom with a $\pi$-pulse
is transferred to the first resonator. This operation is repeated $N$ times, which takes a time $\tau_{1} \propto N$, see Fig.~\SubFig{Fig:NooNF}{b}. We then apply a $\pi/2$-pulse ($\hat{\pi}_{1/2}$) on $\text{Q}$ to generate the following superposition state:
\begin{equation}
\ket{\psi(\tau_{1})}=\left(\1_{R_{1}}\otimes \hat{\pi}_{1/2} \otimes \1_{R_{2}}\right)\ket{Ng0}=\frac{1}{\sqrt{2}}\left(\ket{Ng0}+\ket{Ne0}\right),
\end{equation}
with $\1_{R_{k}}$ the identity matrix for the $k$-th resonator. Finally, the system is let to evolve according to Hamiltonian~(\ref{Eq:Hfull}) by setting the two resonators on resonance, and the qubit to the idling frequency given by Eq.~\eqref{Eq:Idle_Flux}. After an interaction time $\tau_{2}=\pi/[2\left(\kappa - g_{1}g_{2}/\Delta\right)]$, where $(\kappa - g_{1}g_{2}/\Delta)$ is the effective coupling between the resonators when the qubit is in the excited state, the component $\ket{Ne0}$ of the state has changed into $\ket{0eN}$. Yet the component $\ket{Ng0}$ with the qubit in the ground remains unchanged, since the information can flow between the resonators only when the qubit is in state $\ket{e}$. This particular dynamics is a consequence of the state-dependence of  the idling point, and the system ends in state
\begin{align}
    \ket{\psi(\tau_{2})}=\frac{1}{\sqrt{2}}\left(\ket{Ng0}+e^{i\theta}\ket{0eN}\right),
    \label{eq:NO1}
\end{align}
where $e^{i\theta}$ is a dynamical relative phase. Then, we apply a second $\pi/2$-pulse on the qubit, sending the system into
\begin{align}
    \ket{\psi_{\mathrm{out}}}=\frac{1}{2}(\ket{Ng0}+e^{i\theta}\ket{0gN}) + \frac{1}{2}(\ket{Ne0}-e^{i\theta}\ket{0eN}).
\end{align}
Finally we perform a measurement of the qubit state, given the positive operator valued measures $\{M_{0}=\1_{R_{1}}\otimes \ketbra{g}{g} \otimes \1_{R_{2}}, M_{1}=\1_{R_{1}}\otimes \ketbra{e}{e} \otimes \1_{R_{2}}\}$ such that
\begin{equation}
\ket{\Psi_{i}}=\frac{M_{i}\ket{\psi_{\mathrm{out}}}}{\sqrt{\bra{\psi_{\mathrm{out}}}M_{i}\ket{\psi_{\mathrm{out}}}}}, \ \ \ i\in \{0,1\}.
 \end{equation}
The possible outputs of this measurement are, with equal probability, the following system states
\begin{align}
    &\ket{\Psi_{0}}=\frac{1}{\sqrt{2}}(\ket{Ng0}+e^{i\theta}\ket{0gN}),\\
    &\ket{\Psi_{1}}=\frac{1}{\sqrt{2}}(\ket{Ne0}-e^{i\theta}\ket{0eN}).
\end{align}
Therefore, by monitoring the measured qubit state, $\ket{g}$ or $\ket{e}$, we obtain a NOON state, with a relative phase which can be eliminated~\cite{Hofheinz_2009}.  Hence, our protocol is able to deterministically generate maximally entangled states between the resonators.

The high fidelity of the NOON states obtained with this protocol is illustrated in Fig.~\SubFig{Fig:NooNF}{a}, where the scaling of its fidelity with the photon number $N$ is plotted. The fidelity reduces as $N$ increases,  due to the superradiant-like enhancement in the effective coupling. Indeed, when an increasing number $N$ of photons are present in the modes, the validity of dispersive regime approximation is affected, so the dynamics deviates from the ideal one given by the effective Hamiltonian~\eqref{Eq:Heff}. This can be compensated by reducing the qubit-resonator $g_{k}$, reentering the dispersive regime, as can be observed in Fig.~\SubFig{Fig:NooNF}{a}. For the parameters under consideration, our protocol is able to generated post-selection NOON states with a fidelity over $99\%$.

The advantage of our protocol with respect to the previous proposal in Ref.~\cite{Strauch:10} can be found in the number of steps required to generated a NOON state with $N$ excitations. Indeed, as depicted in Fig.~\SubFig{Fig:NooNF}{b}, only $2N+4$ steps are needed to achieve a $N$-excitation state: $2N$ to generate state $\ket{Ng0}$ and $4$ steps in the $\hat{\pi}_{1/2}\rightarrow\text{Interaction time}\rightarrow\hat{\pi}_{1/2}\rightarrow\text{Measurement}$ sequence. Differently, the protocol proposed in Ref.~\cite{Strauch:10} require $4N$ steps, which makes it more prone to errors. The difference in step numbers is particularly advantageous when the number of excitations grows.



\emph{Atoms in optical cavities.---} Our study opens new prospects for a broad class of systems where qubits and resonators are coupled, such as crossed optical cavities ~\cite{Brekenfeld_2020,Villa_2018}. For these setups, the collateral interaction between the modes is derived through the classical Hamitonian $H=(\vec{p}-q\vec{A})^{2}/(2m)+qU$ of a particle interacting with a radiation field, where $\vec{p}$, $q$ and $m$ are the momentum, net charge and mass of the charged particle -- here a valence electron. $\vec{A}$ and $U$ represent the vector and scalar potential of the field, respectively. The collateral interaction here stems from the term $ q^{2}\vec{A}^{2}/(2m)$, as already identified by Fermi~\cite{Fermi_32}, yet then neglected in optical cavity experiments. Moving to SI units and quantizing the vector potential $\vec{A}$, we obtain 
\begin{align}
\vec{A}=\sum_{k,\lambda}\left(\frac{\hbar}{2\varepsilon_{0}\omega_{k}V}\right)^{1/2}\left(\hat{a}_{k\lambda}(t)e^{i\vec{k}\cdot \vec{r}}\hat{e}_{k\lambda}+\hat{a}^{\dagger}_{k\lambda}(t)e^{-i\vec{k}\cdot \vec{r}}\hat{e}^{*}_{k\lambda}\right).
\end{align}
This leads to the interaction between two modes of frequency $\omega_1$ and $\omega_2$~\cite{Fermi_32}: 
    $\kappa_{c}=\frac{q^{2}}{4m\varepsilon_{0}V\sqrt{\omega_{1}\omega_{2}}}$, 
with $\varepsilon_{0}$ the vacuum permittivity and $V$ the volume in which the vector potential is confined. 

Let us now consider that these two cavities are 
resonant with the transition of the atom which mediates the interaction between them (frequency $\omega_a=\omega_1=\omega_2$). The atom-mode coupling is then given by $g=d\sqrt{\omega_a/2\varepsilon_0 \hbar V}$, with $d$ the electric dipole moment of the transition. The transition linewidth being given by $\Gamma=\omega_a^3d^2/3\pi\varepsilon_0\hbar c^3$, with $c$ the light velocity, we obtain
$g/\kappa_c=m\varepsilon_0\sqrt{24\pi Vc^3\Gamma}/q^2$.
For the most recent setups of crossed optical cavities~\cite{Brekenfeld_2020} with a mode volume $V\sim 10^4\mu$m$^2$, the collateral term is several orders of magnitude smaller than the coupling $g$ for a MHz transition linewidth. Nevertheless, for a mHz linewidth such as those explored for clocks~\cite{Norcia2016}, we obtain a ratio $g/\kappa_c\approx 40$. Then, as in the case superconducting circuit, the collateral coupling may have a drastic effect on the dynamics, allowing for an idling point and thus an additional mechanism to operate quantum gates, for example. Indeed, while parasitic coupling between resonators in superconducting circuits do allow to achieve idling points, such couplings are in principle absent from optical cavities setups.


\emph{Conclusion and perspectives.---} We have thus discussed how a collateral interaction arises when a qubit is used to mediate an interaction between the resonators of a superconducting circuit, as one transforms the Lagrangian of the system into a Hamiltonian to obtain dynamical equations. Despite the interaction has an amplitude much weaker than the direct resonator-qubit coupling, it is able to induce an idling point between the resonators: Even for resonant resonators, it prevents the flow of information between them. However, the state-dependent nature of the idling point actually offers a new control parameter for the system, which allowed us to propose a scheme for a fast and high fidelity generation of maximally entangled NOON states between the superconducting resonators. Equally present in crossed optical cavities interacting through cold atoms, the effect could provide a new control parameter to operate quantum gates in these setups as well.


\begin{acknowledgments}
The authors acknowledge the financial support of the São Paulo Research Foundation (FAPESP) (Grants No. 2018/15554-5, 2019/22685-1, 2019/13143-0, 2021/10224-0 and 2022/12382-4) and of the Brazilian CNPq (Conselho Nacional de Desenvolvimento Científico e Tecnológico), Grants No. 130267/2022-8, 313886/2020-2, 465469/2014-0, and 311612/2021-0.
\end{acknowledgments}






\bibliography{mybib-URL}

\onecolumngrid
\newpage

\begin{center}
	{\large{ {\bf Supplemental Material for: \\ Collateral coupling between superconducting resonators: Fast and high fidelity \\generation of qudit-qudit entanglement}}}

\vskip0.5\baselineskip{Pedro Rosario$^{1}$, Alan C. Santos~\orcidlink{0000-0002-6989-7958}$^{1,\ast}$, C. J. Villas-Boas~\orcidlink{0000-0001-5622-786X}$^{1}$ and R. Bachelard~\orcidlink{0000-0002-6026-509X}$^{1,2,\dagger}$}

\vskip0.5\baselineskip{{\em$^{1}$Departamento de Física, Universidade Federal de São Carlos,\\ Rodovia Washington Luís, km 235 - SP-310, 13565-905 São Carlos, SP, Brazil}\\
	{\em $^{2}$Universit\'e C\^ote d'Azur, CNRS, Institut de Physique de Nice, 06560 Valbonne, France}
}

\vskip0.5\baselineskip{$^{\color{blue}\ast}$ac\_santos@df.ufscar.br, ~~~ $^{\color{blue}\dagger}$romain@ufscar.br}
\end{center}

\appendix

\setcounter{equation}{0}
\setcounter{figure}{0}
\setcounter{table}{0}

\renewcommand{\theequation}{S\arabic{equation}}
\renewcommand{\thefigure}{S\arabic{figure}}
\renewcommand{\bibnumfmt}[1]{[S#1]}
\renewcommand{\citenumfont}[1]{S#1}

\twocolumngrid

\section{System description}

\begin{figure}[b!]
	\includegraphics[width=0.8\columnwidth]{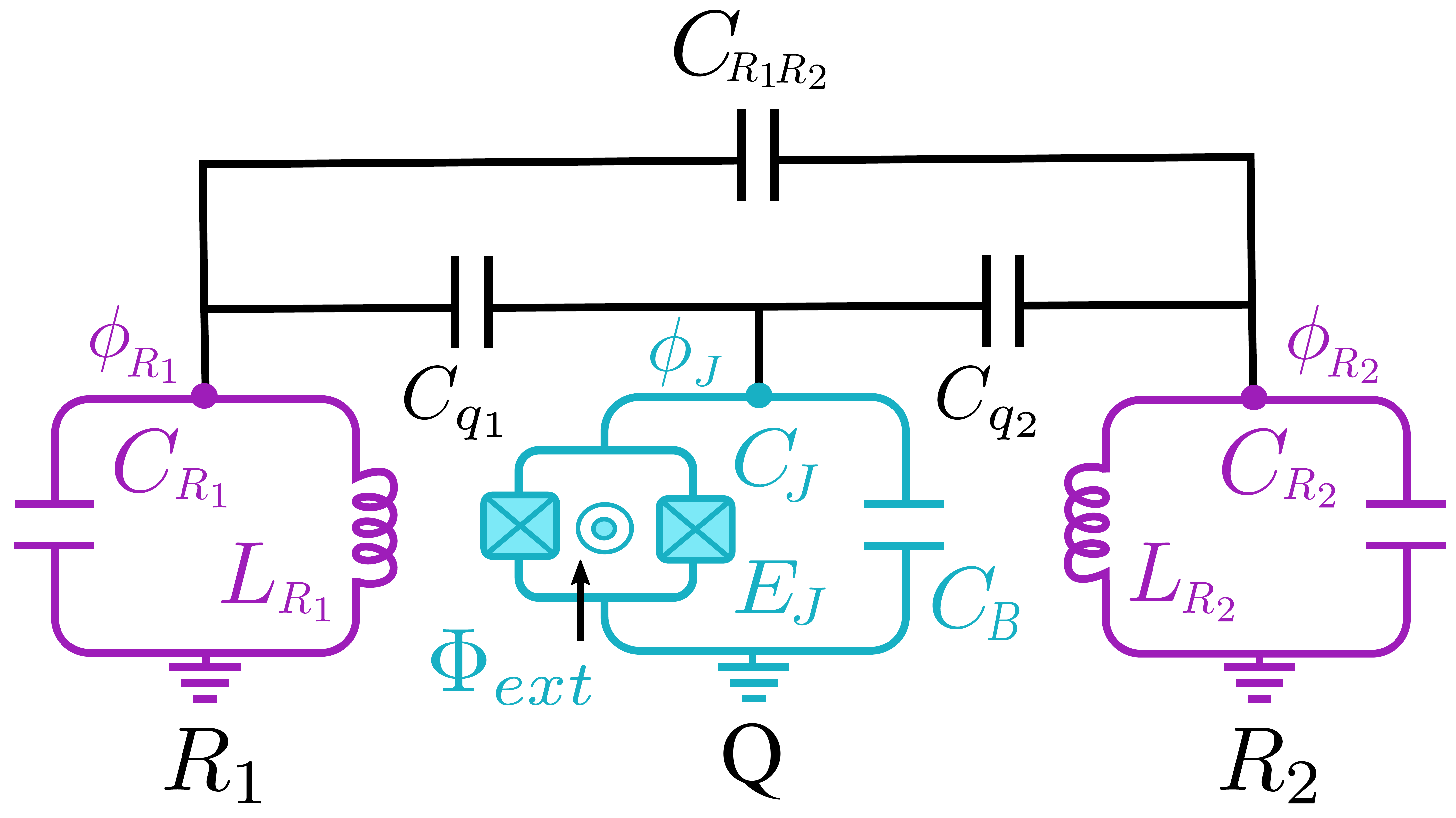}
	\caption{Schematic representation considering the direct parasitic  capacitive interaction $C_{R_{1}R_{2}}$.}
	\label{Fig:Total_system}
\end{figure}

Consider the superconducting circuit of two resonators and a single transmon qubit where capacitive parasitic interaction introduce a new capacitance $C_{R_{1}R_{2}}$ between the resonators, as sketched in Fig.~\ref{Fig:Total_system}. The Lagrangian of the system has the general form
\begin{align}
    L=\frac{1}{2}\vec{\dot{\phi}}^{T}\hat{C}\vec{\dot{\phi}}-U(\phi_{J},\phi_{R_{1}},\phi_{R_{2}}), \label{eq:Lag2}
\end{align}
where $\vec{\dot{\phi}}^{T}=(\dot{\phi_{J}},\dot{\phi}_{R_{1}},\dot{\phi}_{R_{2}})$ and $\hat{C}$ is the capacitance matrix. Then, the capacitance matrix can be derived from Eq.\eqref{eq:Lag2}, using the conjugate momentum of the node flux  $q_{k} = \partial L(\vec{\phi},\vec{\dot{\phi}}) / \partial \mathbin{\textcolor{black}{\dot{\phi_{k}}}}$, and one obtains
\begin{align}
	\hat{C}=\begin{pmatrix}
		C_{\mathrm{T}} +C_{q_{1}}+C_{q_{2}} & -C_{q_{1}} & -C_{q_{2}}\\
		-C_{q_{1}} &C_{R_{1}}+C_{q_{1}}+C_{R_{1}R_{2}}&-C_{R_{1}R_{2}}\\
		-C_{q_{2}} &-C_{R_{1}R_{2}} & C_{R_{2}}+C_{q_{2}}+C_{R_{1}R_{2}}
	\end{pmatrix} .
 \label{eq:C_App}
\end{align}

Switching to the Hamiltonian approach we get:
\begin{align}
	H=\frac{1}{2}\vec{q}^{T}\hat{C}^{-1}\vec{q} + U(\phi_{J},\phi_{R_{1}},\phi_{R_{2}})
 \label{eq:H_App}
\end{align}
where $U(\vec{\phi})=\frac{1}{2L_{R_{1}}}\phi^{2}_{R_{1}} +\frac{1}{2L_{R_{2}}}\phi^{2}_{R_{2}}-\tilde{E}_{J}\cos(\phi_{J} / \phi_{0})$, $\vec{q}^{T} = (q_{J},q_{R_{1}},q_{R_{2}})$, and $\hat{C}^{-1}$ is the inverse of the capacitance matrix, which induces the collateral interaction between the resonators. Indeed, if one then assumes that $C_{\mathrm{T}} \gg C_{q_{k}}$ and $C_{R_{k}} \gg C_{q_{k}}$, one can approximate the terms $C_{\mathrm{T}} +C_{q_{1}}+C_{q_{2}} \approx C_{\mathrm{T}}$ and $C_{R_{k}}+C_{q_{k}} \approx C_{R_{k}}$ in Eq.~\eqref{eq:C_App}, to obtain the following inverse capacitance matrix
\begin{align}
	\hat{C}^{-1}\approx \begin{pmatrix}
		\frac{1}{C_{T}} & \frac{C_{q_{1}}C_{R_{2}}}{\text{Det}(\hat{C})}  & \frac{C_{q_{2}}C_{R_{1}}}{\text{Det}(\hat{C})}  \\
		\frac{C_{q_{1}}C_{R_{2}}}{\text{Det}(\hat{C})} &\frac{C_{T}(C_{R_{2}}+C_{R_{1}R_{2}})}{\text{Det}(\hat{C})} &\frac{C_{q_{1}}C_{q_{2}}+C_{R_{1}R_{2}}C_{T}}{\text{Det}(\hat{C})} \\
		\frac{C_{q_{2}}C_{R_{1}}}{\text{Det}(\hat{C})} &\frac{C_{q_{1}}C_{q_{2}}+C_{R_{1}R_{2}}C_{T}}{\text{Det}(\hat{C})} & \frac{C_{T}(C_{R_{1}}+C_{R_{1}R_{2}})}{\text{Det}(\hat{C})} 
	\end{pmatrix} .
\end{align}
where $\text{Det}(\hat{C})\approx C_{T}(C_{R_{1}}C_{R_{2}}+C_{R_{1}R_{2}}(C_{R_{1}}+C_{R_{2}}))$. The charge-charge capacitive coupling between the resonators reads
\begin{align}
    \frac{1}{\text{Det}(\hat{C})}(C_{q_{1}}C_{q_{2}}+C_{R_{1}R_{2}}C_{T})q_{R_{1}}q_{R_{2}}
    \label{eq:qqc}
\end{align}
Note that, in the regime where the physical distance between the resonators is large enough, the direct capacitive coupling represented in Fig. \ref{Fig:Total_system} vanishes ($C_{R_{1}C_{R_{2}}}\rightarrow 0 $) and the resonators coupling in Eq. \eqref{eq:qqc} has only the collateral contribution due to the capacitance network
\begin{align}
   \frac{C_{q_{1}}C_{q_{2}}}{C_{R_{1}}C_{R_{2}}C_{\mathrm{T}}}q_{R_{1}}q_{R_{2}}
\end{align}
In our system description, we are only interested in study the effect of the collateral interaction to the quantum dynamics, therefore we do not consider the direct parasitic coupling ($C_{R_{1}C_{R_{2}}}\rightarrow 0 $) and the inverse capacitance matrix takes the form
\begin{align}
	\hat{C}^{-1}\approx \begin{pmatrix}
		\frac{1}{C_{T}} & \frac{C_{q_{1}}}{C_{R_{1}}C_{T}}  & \frac{C_{q_{2}}}{C_{R_{2}}C_{T}}  \\
		\frac{C_{q_{1}}}{C_{R_{1}}C_{T}} &\frac{1}{C_{R_{1}}} &\frac{C_{q_{1}}C_{q_{2}}}{C_{R_{1}}C_{R_{2}}C_{T}} \\
		\frac{C_{q_{2}}}{C_{R_{2}}C_{T}} &\frac{C_{q_{1}}C_{q_{2}}}{C_{R_{1}}C_{R_{2}}C_{T}} & \frac{1}{C_{R_{2}}} 
	\end{pmatrix} .
\end{align}
Then, Hamiltonian~\eqref{eq:H_App} reads
\begin{align}
	H&\approx\frac{q^{2}_{J}}{2C_{\mathrm{T}}}- \tilde{E}_{J}\cos(\frac{\phi_{J}}{\Phi_{0}})
	+ \frac{q^{2}_{R_{1}}}{2C_{R_{1}}} + \frac{\phi^{2}_{R_{1}}}{2L_{R_{1}}} 
	+ \frac{q^{2}_{R_{2}}}{2C_{R_{2}}} + \frac{\phi^{2}_{R_{2}}}{2L_{R_{2}}} \nonumber \\
	&+ \frac{C_{q_{1}}}{C_{R_{1}}C_{\mathrm{T}}}q_{J}q_{R_{1}}+ \frac{C_{q_{2}}}{C_{R_{2}}C_{\mathrm{T}}}q_{J}q_{R_{2}} + \frac{C_{q_{1}}C_{q_{2}}}{C_{R_{1}}C_{R_{2}}C_{\mathrm{T}}}q_{R_{1}}q_{R_{2}},
\end{align}
with the collateral term now becomes explicit. 
The quantization of the circuit is done through the canonical quantization that consists in turning the functions $q$ and $\phi$ for their respective operators $\hat{q}$ and $\hat{\phi}$, along with the second quantization approach where the creation and annihilation operators for the resonators are introduced
\begin{align}
	\hat{\phi}_{R_{k}}=\left(\frac{\hbar Z_{R_{k}}}{2}\right)^{1/2}(\hat{a_{k}}+\hat{a_{k}}^{\dagger}), ~~ \hat{q}_{R_{k}}=i\left(\frac{\hbar}{2Z_{R_{k}}}\right)^{1/2}(\hat{a_{k}}^{\dagger}-\hat{a_{k}}),
\end{align}
with $Z_{R_{k}}=\sqrt{L_{R_{k}}/C_{R_{k}}}$ the impedance of each resonator. The transmon variables are given by
\begin{align}
	\frac{\hat{\phi}_J}{\Phi_0}=\frac{1}{\sqrt{2}}\left(\frac{8E_C}{\tilde{E}_{J}}\right)^{1/4}(\hat{b}+\hat{b}^{\dagger}), ~~ \hat{q}_{J}=-\frac{2ie}{\sqrt{2}}\left(\frac{\tilde{E}_{J}}{8E_C}\right)^{1/4}(\hat{b}-\hat{b}^{\dagger}) ,
\end{align}
with $E_C=e^{2}/2C_{\mathrm{T}}$. Since the Josephson energy $\tilde{E}_{J}$ is much larger than the capacitance energy $E_{C}$, one can expand and approximate the contribution of the Josephson junction to the Hamiltonian as~\cite{Rasmussen:21}
\begin{align}
	\tilde{E}_{J}\cos(\frac{\hat{\phi}_{J}}{\Phi_{0}}) \approx \tilde{E}_{J} \left[ \1 - \frac{1}{2}\left(\frac{\hat{\phi}_{J}}{\Phi_{0}}\right)^2 + \frac{1}{4!}\left(\frac{\hat{\phi}_{J}}{\Phi_{0}}\right)^4  \right] . \label{Eq:EnergyExpansion}
\end{align}
Eventually, one gets
\begin{align}
	H&=\hbar\omega_q \left(\hat{b}^{\dagger}\hat{b}+\frac{\1}{2}\right)-\frac{E_C}{2}\hat{b}^{\dagger}\hat{b}^{\dagger}\hat{b}\hat{b}+\hbar\omega_{R_{1}}\left(\hat{a}_{1}^{\dagger}\hat{a}_{1}+\frac{\1}{2}\right)\nonumber\\
&+\hbar\omega_{R_{2}}\left(\hat{a}_{2}^{\dagger}\hat{a}_{2}+\frac{\1}{2}\right) +\hbar g_{1}(\hat{b}-\hat{b}^{\dagger})(\hat{a}_{1}^{\dagger}-\hat{a}_{1}) \nonumber\\
	&+\hbar g_{2}(\hat{b}-\hat{b}^{\dagger})(\hat{a}_{2}^{\dagger}-\hat{a}_{2}) - \hbar\kappa(\hat{a}_{1}^{\dagger}-\hat{a}_{1})(\hat{a}_{2}^{\dagger}-\hat{a}_{2}) , \label{Eq:Ap_Hfull}
\end{align} 
where $\hbar \omega_q = \sqrt{8E_C \mathbin{\textcolor{black}{\tilde{E}_{J}}}}-E_C$ and $\omega_{R_{k}}=1 / \sqrt{C_{R_{k}} L_{R_{k}}}$ are the qubit and resonators frequencies, respectively. The coupling strengths read
\begin{subequations}
\label{Eq:Parameters}
\begin{align}
	g_{k} &= - \frac{1}{\hbar } \frac{C_{q_{k}}}{\sqrt{ C_{\mathrm{T}} C_{R_{k}}}} \left(2E_C\tilde{E}_{J}\epsilon_{R_{k}}^2\right)^{\frac{1}{4}}, \\
	\kappa &= \frac{C_{q_{1}}C_{q_{2}}}{C_{\mathrm{T}}\sqrt{C_{R_{1}}C_{R_{2}}}}\frac{\sqrt{\epsilon_{R_{2}}\epsilon_{R_{1}}}}{\hbar} ,
\end{align}
\end{subequations}
with $\epsilon_{R_{k}} = \hbar \omega_{R_{k}}/2$ the zero-point energy of the $k$-th resonator. In general the energy in the superconducting device has the form ~\cite{Blais_2021_re}
\begin{align}
    \tilde{E}_{J}=(E_{J_{1}}+E_{J_{2}})\cos{\left(\frac{\Phi_{\mathrm{ext}}}{2\Phi_{0}}\right)}\sqrt{1+d^{2}\tan^{2}{\left(\frac{\Phi_{\mathrm{ext}}}{2\Phi_{0}}\right)}}
\end{align}
with $d=\frac{E_{J_{1}}-E_{J_{2}}}{E_{J_{1}}+E_{J_{2}}}$ the asymmetry parameter. Considering two Josephson junctions with the same associated energy ($d=0$), the energy rewrites $\tilde{E}_{J}=\tilde{E}_{J}(\Phi_{\mathrm{ext}}) = 2 E_{J} \cos(\Phi_{\mathrm{ext}}/2\Phi_{0})$. Eq. \eqref{Eq:Hfull} in the main text is obtained from Eq.~\eqref{Eq:Ap_Hfull} using the second level approximation $\hat{b}^{\dagger}:=\hat{\sigma}^{+}$, $\hat{b}:=\hat{\sigma}^{-}$ and neglecting the non-conserving terms of the form $\hat{b}^{\dagger}\hat{a_{j}}^{\dagger}$, $\hat{b}\hat{a_{j}}$, $\hat{a_{i}}\hat{a_{j}}$ and $\hat{a_{i}}^{\dagger}\hat{a_{j}}^{\dagger}$ with $j\neq i \in \{1,2\}$.


    \section{Effective Hamiltonian and idling point}
\label{sec:EffectiveH}

Consider resonators on resonance $\omega_{R_{1}}=\omega_{R_{2}}=\omega_{R}$, then the detuning of the qubit is the same for both resonators $\Delta=\omega_{q}-\omega_{R}$. The interaction Hamiltonian  is given by the transformation $\textcolor{black}{H_{\mathrm{in}}(t)=e^{iH_{\mathrm{fs}}t/\hbar}H_{\mathrm{I}}e^{-iH_{\mathrm{fs}}t/\hbar}}$, where $H_{\mathrm{fs}}=\hbar \omega_q \hat{\sigma}^{+}\hat{\sigma}^{-}+\hbar\omega_{r_{1}}\hat{a}_{1}^{\dagger}\hat{a}_{1} + \hbar\omega_{r_{2}}\hat{a}_{2}^{\dagger}\hat{a}_{2}$ describes the local contribution of each element in the system.  $H_{\mathrm{I}}=\hbar g_1 (\hat{\sigma}^{-}\hat{a}_{1}^{\dagger}+\hat{\sigma}^{+}\hat{a}_{1})+ \hbar g_2 (\hat{\sigma}^{-}\hat{a}_{2}^{\dagger}+\hat{\sigma}^{+}\hat{a}_{2})+\hbar \kappa (\hat{a}_{1}^{\dagger}\hat{a}_{2}+\hat{a}_{1}\hat{a}_{2}^{\dagger})$ represents the interactions transmon-resonator and resonator-resonator, therefore it also encodes the collateral contribution to the dynamics. After some algebra we obtain
\begin{align}
   \nonumber H_{\mathrm{in}}(t)&=\hbar g_{1}(e^{i\Delta t}\sigma^{+}\hat{a_{1}}+e^{-i\Delta t}\sigma^{-}\hat{a_{1}}^{\dagger})\\
    &+\hbar g_{2}(e^{i\Delta t}\sigma^{+}\hat{a_{2}}+e^{-i\Delta t}\sigma^{-}\hat{a_{2}}^{\dagger})+\hbar\kappa(\hat{a_{1}}^{\dagger}\hat{a_{2}}+\hat{a_{1}}\hat{a_{2}}^{\dagger}).
\end{align}
The effective Hamiltonian is obtained by eliminating the fast rotating terms (rotating wave approximation) assuming that $|\Delta|\gg g_{k}$:
\begin{align}
    \nonumber H_{\mathrm{eff}}(t)&\approx \frac{1}{i\hbar} H_{\mathrm{in}}(t)\int_{0}^{t}H_{\mathrm{in}}(t')dt'\\
    &\nonumber=\hbar\kappa(\hat{a_{1}}^{\dagger}\hat{a_{2}}+\hat{a_{1}}\hat{a_{2}}^{\dagger})\\
    &+\nonumber\frac{\hbar g^{2}_{1}}{\Delta}\left(\hat{\sigma}^{+}\hat{\sigma}^{-}\hat{a_{1}}\hat{a_{1}}^{\dagger}-\hat{\sigma}^{-}\hat{\sigma}^{+}\hat{a_{1}}^{\dagger}\hat{a_{1}}\right)\\
    &\nonumber+\frac{\hbar g_{1}g_{2}}{\Delta}\left(\hat{\sigma}^{+}\hat{\sigma}^{-}\hat{a_{1}}\hat{a_{2}}^{\dagger}-\hat{\sigma}^{-}\hat{\sigma}^{+}\hat{a_{1}}^{\dagger}\hat{a_{2}}\right)\\
    &\nonumber+\frac{\hbar g^{2}_{2}}{\Delta}\left(\hat{\sigma}^{+}\hat{\sigma}^{-}\hat{a_{2}}\hat{a_{2}}^{\dagger}-\hat{\sigma}^{-}\hat{\sigma}^{+}\hat{a_{2}}^{\dagger}\hat{a_{2}}\right)\\
    &+ \frac{\hbar g_{1}g_{2}}{\Delta}\left(\hat{\sigma}^{+}\hat{\sigma}^{-}\hat{a_{2}}\hat{a_{1}}^{\dagger}-\hat{\sigma}^{-}\hat{\sigma}^{+}\hat{a_{2}}^{\dagger}\hat{a_{1}}\right).
\end{align}
which can be rewritten as Eq.~\ref{Eq:Heff} from the main text.
This expression can in turn be used to determine the idling point, where the resonators stop exchanging excitations. This occurs when the effective coupling between the resonators is null with the qubit in state $\ket{g}$:
\begin{align}
   \frac{ g^{2}_{k}}{\Delta}+\kappa=0,
\end{align}
where we have used that $g_{1}=g_{2}=g_{k}$. This leads to the following conditions on the external flux for the  dynamics to be static: 
\begin{align}
    \nonumber&0\overset{1}{=}\frac{1}{\hbar}\frac{C^{2}_{q_{k}}}{ C_{\mathrm{T}} C_{R_{k}}}\frac{ \left(2E_C\tilde{E}_{J}\epsilon_{R_{k}}^2\right)^{\frac{1}{2}}}{ \sqrt{8E_C \mathbin{\textcolor{black}{\tilde{E}_{J}}}}-E_C -\hbar \omega_{R_{k}}}+\kappa\\
    &\nonumber0\overset{2}{=}\frac{1}{\hbar}\frac{C^{2}_{q_{k}}}{ C_{\mathrm{T}} C_{R_{k}}} \left(2E_C\tilde{E}_{J}\epsilon_{R_{k}}^2\right)^{\frac{1}{2}}+\kappa\left(\sqrt{8E_C \mathbin{\textcolor{black}{\tilde{E}_{J}}}}-E_C -\hbar \omega_{R_{k}}\right)\\
    &0\overset{3}{=}\frac{2\kappa C_{T}C_{R_{k}}(E_{C}+\hbar\omega_{R_{k}})}{\sqrt{2E_{C}}(C^{2}_{q_{k}}\omega_{R_{k}}+4\kappa C_{T}C_{R_{k}})}-\sqrt{\tilde{E}_{J}}.
\end{align}
Using the last equation, we obtain expression \eqref{Eq:Idle_Flux} presented in the main text. 

\begin{figure}[th!]
	\includegraphics[width=0.7\columnwidth]{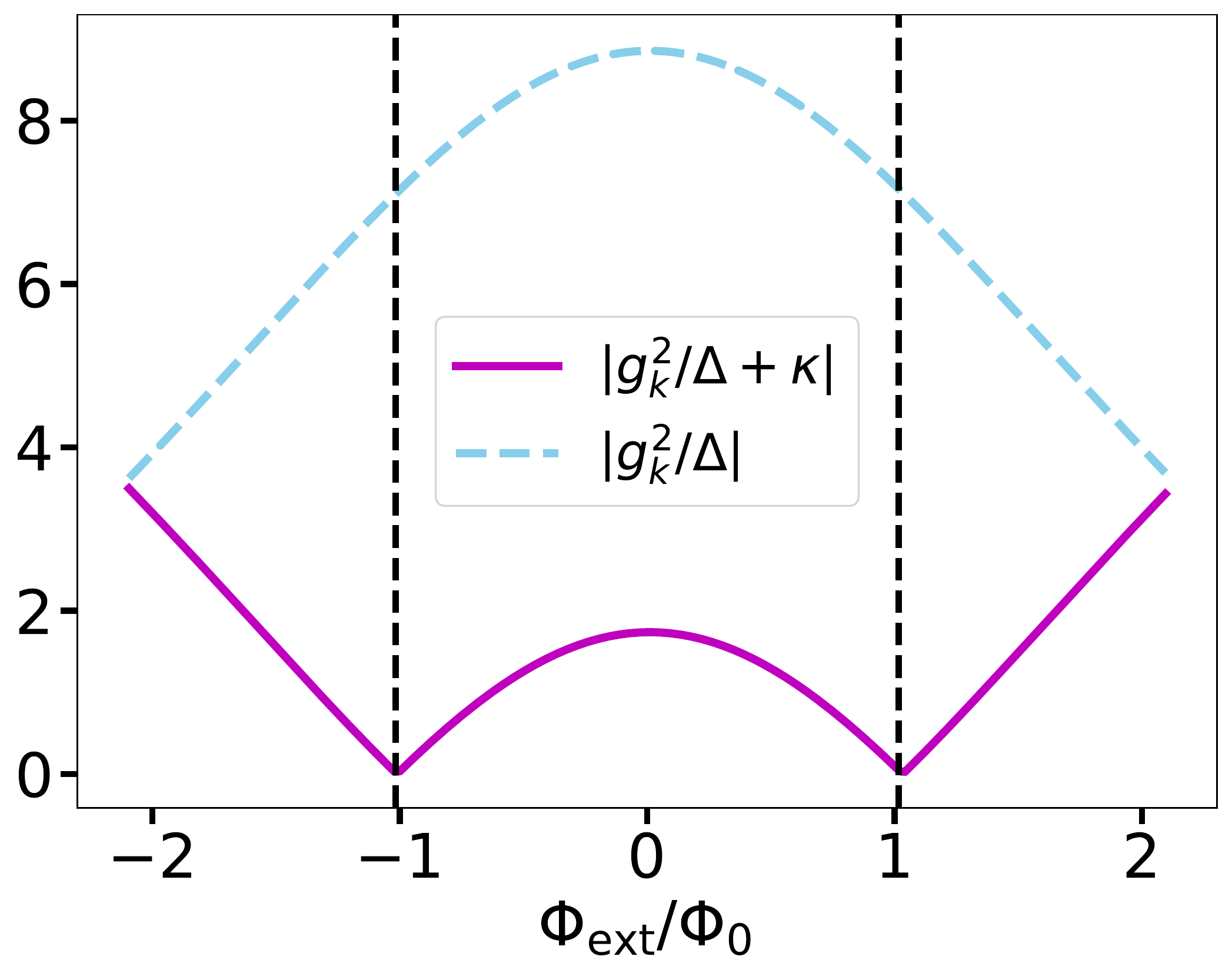}
	\caption{Absolute value of the effective coupling as a function of the external flux, with and without $\kappa$. Vertical lines corresponds to the analytical formula~\eqref{Eq:Idle_Flux}.}
	\label{Fig:EffectiveC}
\end{figure}

In Fig.~\ref{Fig:EffectiveC} we present the effective coupling as a function of the external flux, with and without collateral interaction: This illustrates the necessity of a finite collateral coupling $\kappa$ to generate the idling point.

\end{document}